\input amstex      
\loadeusm      
\documentstyle{amsppt}       
\refstyle{B}
\magnification=\magstephalf
\document   
\magnification=\magstephalf
\NoBlackBoxes      
\redefine\a{\alpha}      
\redefine\b{\beta}      
\redefine\d{\delta}      
\redefine\g{\gamma}      

\redefine\k{\kappa}      
\redefine\la{\lambda}      
\redefine\G{\Gamma}

\topmatter
\title
 Multipole moments in Kaluza-Klein   theories
\endtitle
\author Ewa Czuchry {\rm and} Wojciech Kopczy\'nski \endauthor
\affil
Instytut Fizyki Teoretycznej, Uniwersytet Warszawski, \\ul. Ho\.za 69, 00-681 Warszawa, Poland 
\endaffil
\date August  1997 \enddate
\abstract
{This paper contains discussion of the problem of motion of extended
i.e. non point test bodies in multidimensional space. Extended bodies
are described in terms of so called multipole moments. Using
approximated form of equations of motion for extended bodies
deviation from geodesic motion is derived. Results are applied to
special form of space-time. 
\par
}
\noindent
PACS number(s): 04.25.-g, 04.50.+h, 11.25.Mj
\endabstract
\endtopmatter   
 
\head 1. Introduction\endhead 

Kaluza-Klein type theories have regained a lot of
interest \cite{1} over recent years. 
It was mainly due to the construction of gauge theories that
the old idea of adding a
compactified dimension introduced by Kaluza and Klein \cite{2, 3} has been
developed and utilised in a variety of theories. All of them are based
on an assumption that the standard four-dimensional world is a subset
of a higher--dimensional space time, with extra space dimensions
dynamically compactified.

However these types of Kaluza--Klein theories leave an intriguing
question unanswered: where the distinction between the observed  
space--like dimensions and other compactified ones comes from. One
possible solution to this problem was given by Kopczy\'nski \cite{4}.
His idea was to assume that this is the form of matter that gives us
this distinction. Constituents of matter would be not points in a
$m$-dimensional space-time but rather $(k{-}1)$-dimensional objects ---
strings, membranes etc. There exists a cosmological solution \cite{5}
such that we have expanding $m{-}k{+}1$ space like dimensions and
contracting $k$ dimensions filled with matter.

Then the question arises whether that extendness in additional space
dimensions has any influence on macroscopic properties of matter.
The purpose of this paper is  to look at behaviour of test bodies ---
extended in compactified dimensions --- moving in the observed
$(m{-}k{+}1)$-dimensional space-time. We will show that the non point
structure of these objects modifies their equations of motion, which
results in deviations from the geodesic motion. Furthermore, these
deviations depend on the topological structure of the compactified
space.

It is highly nontrivial to obtain any exact solution of such a
problem.  Therefore we will follow another way. We will describe
extended bodies in terms of multipole moments and write equation of
motion in that approximate way \cite{6}.

The structure of that paper is following. In section 2. we briefly
discuss equations of motion for extended bodies and introduce some
basic mathematical tools. In section 3. we define the problem and 
discuss initial assumptions that will build the model of space-time 
to be considered 
afterwards. In sections 4. and 5. we derive equation of
motion in dipole and multipole approximation. In section 6. we estimate 
form of multipole moments applying results from previous sections to 
a special form of space-time.

\head 2. Equations of motion for extended bodies \endhead

In order to describe extended bodies first we choose some unique world line
laying inside the world tube of the body that represents its dynamical
properties. Although in General Relativity there are many existing
possible unique choices we follow the ideas of Dixon \cite{6} and
Beiglb\"{o}ck \cite{7}. Thus we imply that the best such a line,
called the centre-of-mass world line, is the one such that at at every point
of it momentum $p^\k$ and angular momentum $S^{\k\la}$ are perpendicular:
$$p_\k S^{\k\la}=0\;.\tag{1}$$
We parametrise this world line  affinely as $z(s)$. Then we
introduce coordinate system $X^\k$ at the given point z of the
centre-of-mass world line in the way that  any point $x$ of space-time
is given by a  vector:
$$X^\k=-\sigma^\k(z,x)\;,\tag{2}$$
where $\sigma(z,x)$ is the biscalar world-function
\cite{8}. Properties of the world function imply that $X^\k$
is  a
two-point vector defined at points $z$ and
$x$ such that its length is equal do geodesic distance between that
points.

On the centre-of-mass world line we define tensorial multipole moments
of the energy-momentum tensor. Because those moments describe inner
properties of matter we would rather like to separate them from
external space-time properties. As ordinary energy-momentum tensor
describes both internal properties of matter and external geometry we
introduce another quantity: {\it energy-momentum skeleton}
$\hat{T}^{\k\la}$. It is obtained from energy-momentum tensor
$T^{\k\la}$ by subtraction from $T^{\k\la}$ information on geometry of
space-time. Therefore it
describes only properties of the given body and not external
space-time. Itself it is in some sense tensor-valued quantity i.e. a
function on the space-time manifold ${\eusm M}$ whose values at any
point $z$ is a tensor valued distribution on the tangent space
$T_z{\eusm M}$ to ${\eusm M}$. Energy-momentum skeleton
$\hat{T}^{\k\la}$ is symmetric in its indices $\k$ and $\la$ and is
zero unless $z$ lies on the centre-of-mass line.  Relation between
energy-momentum skeleton $\hat{T}^{\k\la}$ and energy-momentum tensor
$T^{\k\la}$ is derived and discussed in \cite{6}. 

Definition of the $2^n$-pole moment at the given point $z$ of the
centre-of-mass world line is following:
$$I^{\k_1\cdots\k_n\la\mu}(z(s))=\int X^{\k_1}\cdots X^{\k_n}      
\hat{T}^{\la\mu}(z,X)\,\text{D}X\; ,\tag{3}$$    
where $\text{D}X$ is a volume element.
The monopole and the dipole moments are the momentum $p^\k$ and the
angular momentum $S^{\k\la}$ respectively.  
 
Equations of motion
express absolute derivatives along the world line $\d p^\la
/\text{d}s$ and $\d S^{\la\mu} /\text{d}s$ in terms of curvature
tensor and higher multipole moments of the body. 
Generally equations of motion in $2^n$-pole approximation are
following \cite{6}: 
$$ 
\align       
\frac{\d}{\text{d}s}p_k&=\frac{1}{2}v^\la S^{\mu\nu}      
R_{\k\la\mu\nu}      
+\frac{1}{2}\sum_{n=2}^{N}\frac{1}{n!}I^{\nu_1\cdots\nu_n\la\mu}      
\nabla_\k g_{\la\mu ,\nu_1\cdots\nu_n}\; ,\tag{4}\\      
\frac{\d}{\text{d}s}S^{\k\la}&=2p^{{[}\k}v^{\la {]}}+      
\sum_{n=1}^{N-1}\frac{1}{n!}g^{\sigma[\k}      
I^{\la]\rho_1\cdots\rho_n\mu\nu}g_{\{\sigma\nu ,\mu\}\rho_1\cdots\rho_n}
\; .\tag{5} \endalign     
$$       
where $g_{\la\mu ,\nu_1\cdots\nu_n}$ denotes $n$th tensorial extension
of metric tensor $g_{\la\mu}$ (see Schouten \cite{9}) and 
$v^\la\equiv \dot{z}^\la$ is a tangent vector to the centre-of-mass world line.

 Neglecting higher than dipole
moments we obtain Mathisson-Papapetrou equations \cite{10, 11}:
 $$      
\align      
\frac{\d }{\text{d}s}&p^k=\frac{1}{2}v^{\la}S^{\mu\nu}      
R_{\;\;\la\mu\nu}^{\k}\; , \tag{6}\\      
\frac{\d}{\text{d}s}&S^{\k\la}=2p^{[\k}v^{\la ]}\; .\tag{7}      
\endalign      
$$  
that describe motion of spinning particles.

\head 3. Space-time structure in Kaluza-Klein theories \endhead

We want to find equations of motion of  $ (k-1)$-dimensional body
moving in $m$-dimensional  space-time $\eusm  M$. We assume its
structure being in the form of $\eusm M_I\times\eusm M_E$ and the
signature of the metric tensor of  $(+,-,\dots,-)$. We introduce
symbols:
$$      
\align      
{\eusm M}_E& \text{ --- }  (m{-}k{+}1)\text{-dimensional  external
space-time},\\ {\eusm M}_I& \text{ --- } (k{-}1)\text{-dimensional
internal space}.     
\endalign      
$$       
Metric is  of  the form:    
$$       
\text{d}s^2=g^2(x^\a)\text{d}s_E^2-f^2(x^\a)\text{d}s_I^2,\;\;\;\;\;      
\a=0,1,...,m{-}1 \; ,\tag{8}       
$$        
where $\text{d}s_E^2$ and $\text{d}s_I^2$ are external and internal
quadratic length elements, written explicitly as:
$$      
\align      
\text{d}s_E^2&=h_{ab}(x^a)\text{d}x^a\text{d}x^b;\;\;\;\;\;a,\;b=0,1,
\ldots,m{-}k \; ,\tag{9}\\      
\text{d}s_I^2&=h_{kl}(x^k)\text{d}x^k\text{d}x^l;\;\;\;\;\;k,\;l=m-k+1,
\ldots,m{-}1 \; .\tag{10}     
\endalign      
$$     
We have introduced conventions of denoting external coordinates by
initial letters of Latin alphabet --- $a,b,c\ldots$ and  internal
coordinates by central ones --- $k,l,m\ldots$. 

We assume, that characteristic length of internal space is 
very small and metric
tensor of external space-time does not depend on internal coordinates.
Moreover we postulate existence of vectors $\xi$ with only internal
components such that Lie derivative of the external metric tensor along the
corresponding vector field vanishes:
 $$\text{{\it\$}}_\xi g_{ab}=0\; .\tag{11}$$       
Using the definition of the Lie derivative we write this expression  
 explicite and obtain the condition:
$$\xi^lg_{ab,l}+g_{al}\xi^l_{,b}+g_{lb}\xi^l_{,a}=0\; .\tag{12}$$      
But from our assumption follows that partial derivatives of internal
components of Killing vector over external indices vanish:
$$\xi^l_{,b}=0,\;\;\;\xi^l_{,a}=0 .\tag{13}$$ 
Thus we obtain
conditions on partial derivatives of components of the metric tensor:
and it follows that external components of metric
tensor $g_{ab}$ depend only on external coordinates:
$$g_{ab}(x^\a)\equiv g(x^a) h_{ab}= g_{ab}(x^a)\; .\tag{14}$$
By imposing some symmetries on internal space --- e.g. by choice of the
Killing vector $\xi^l$ --- we would have obtained next restrictions on
metric tensor $g_{\a\b}$.

Now we assume that internal scale factor $f(x^\a)$ depends only on
external coordinates: $f(x^\a)=f(x^a)\,$. 
Thus coordinates
from $\eusm M_E$ could only introduce additional scale factor at
$\text{d}s_I$, but they cannot change internal geometry  of $\eusm
M_I$.

Summarising the metric on $\eusm M$ is of the form:
$$       
\text{d}s^2=g_{ab}\text{d}x^a\text{d}x^b -f^2(x^a)\text{d}s_I^2   \;.   
\tag{15}
$$       

Internal space $\eusm M_I$ has  symmetries connected with some
symmetry group $G$ acting on  that. So there exist Killing vectors in
amount equal to the number of generators of the group $G$.

Because of lack of internal space boundary we cannot fully use
Beiglb\"ock theorem on uniqueness of choice of the world line of the
centre of mass. From this and due to existence of internal symmetries
we can only specify equivalence class of the centre of mass world
line. Points $x^k$ and $x^{k'}$ of internal space are connected by
equivalence relation if they lay on the same orbit of action of group
$G$:   $[x^k]$. Centre of mass world line in the internal space is
obtained by intersection of the centre of mass world sheet by
surface $x^k=\text{const}$. Thus the centre of mass world sheet is of
the form of $l \times {\eusm M}_I$. Due to assumed symmetries the
world line in external space should not depend on given choice of
coordinates $x^k$.

Thus it makes sense  to use coordinates $v^\a$ of the world line only
in correspondence to external components. Internal components could be
set arbitrary; in calculations we will  put $v^k=0$.

Therefore  we impose conditions on external  components  of momentum and
angular momentum only. The equation: $p_\a S^{\a\b}=0$
takes then the following form:
$$p_a S^{ab}=0\;.\tag{16}$$
For mixed components we  have the constraint:
$$p_k S^{ka}=0\; .\tag{17}$$ 
Thus in 5-dimensional theory, assuming $p_5 \neq 0$,  we obtain on mixed components condition:
$S^{5a}=0$.

We do not impose any restrictions on internal
 components of momentum and angular momentum.

\head 4. Dipole approximation\endhead

We will now determine equations of motion for $(k-1)$-dimensional 
body in the dipole approximation. This approximation should give us
some insight into the form of modified equations  of motion. From
some 4-dimensional consideration \cite{12} we can interpret the dipole
moment as  resulting only from space-like extendness. Thus dipole solution
correspond to extended spinning particles.

Dipole term contains components of the Riemann tensor
$R_{\a\b\g}^{\;\;\;\;\;\;\d}$,  so first we will determine Christoffel
symbols in the coordinates introduced in the previous section. 
We obtain the least amount of independent components of curvature coefficients for the triples of indices: $(abc)$,  $(abk)$, $(ckl)$, $(akl)$, $(cbk$), $(klm)$. Thus we have that Christoffel symbols  are:
$$      
\aligned      
\G^c_{ab}&=\;\frac{1}{2}g^{cd}(\partial_ag_{bd}+\partial_bg_{ad}-      
\partial_dg_{ab})\;,\;\;\\       
\G^c_{kl}&=\;-\frac{1}{2}g^{cd}\partial_dg_{kl}\;,\;\;\\       
\G^k_{ab}&=\;0 \;,\;\;\\ 
\endaligned
\aligned
\G^m_{kl}&=\;\frac{1}{2}g^{mn}(\partial_kg_{ln}+\partial_lg_{kn}-      
\partial_ng_{kl})\;,\;\;\\
\G^k_{al}&=\;\frac{1}{2}g^{km}\partial_ag_{lm} \;,\;\;\\   
\G^c_{kb}&=\;0 \;. 
\endaligned      
$$   
Now we are able to calculate   
 components of curvatures tensor  with internal and external indices.
We divide different sets of indices into such that contain only external components $(abcd)$:
$$      
R_{\;cba}^{d} =\partial_b\G^d_{ac}-      
    \partial_a\G^d_{bc}+\G^e_{ac}\G^d_{eb}-\G^e_{bc}      
    \G^d_{ea} \;,      
$$
and that containing only internal components $(klmn)$:
$$R_{\;mlk}^{n}=\partial_l\G^n_{km}-      
	\partial_k\G^n_{lm}+\G^p_{kl}\G^n_{pm}-      
	\G^p_{lm}\G^d_{pk}\;,
$$
We are left with the rest  mixed components. Almost all of them vanish as follows:
$$
\aligned
R_{\;kba}^{l} =&\;   
       	\frac{1}{2}g^{l}_{\;k}(\partial_b(\frac{1}{f^2}\partial_af^2)      
  	-\partial_a(\frac{1}{f^2}\partial_bf^2))+\frac{1}{4}\frac{1}
{f^4}g^{l}_{\;k}      
 	\partial_af^2\partial_bf^2+\\ &   
    -\frac{1}{4}\frac{1}{f^4}g^{l}_{\;k}      
   \partial_af^2\partial_bf^2=0 \;,\\      
R_{\;lka}^{m} =&\;   
   \partial_a(\frac{h^{mn}}{f^2}f^2      
  (\partial_kh_{nl}+\partial_lh_{nk}-\partial_nh_{kl}))=0 \;,\\      
R_{\;cba}^{k}=&\;0\;,       
\endaligned
$$
except the one type of components of the Riemann tensor with indices $(albk)$:
$$
R_{\;bka}^{l}=      
   \frac{1}{4}g^l_{\;k}\frac{1}{f^4}\partial_a      
   f^2\partial_bf^2-      
  \frac{1}{2}g^l_{\;k}\frac{1}{f^2}\partial_a\partial_bf^2+      
  \frac{1}{2}g^l_{\;k}\frac{1}{f^2}\G^c_{ab}\partial_cf^2\;.      
$$      
Putting obtained elements of  $R^{\a}_{\;\;\b\g\d}$ into equation
(6)  we derive  equations of evolution  along the world line of the
mass centre for external components of momentum $\frac{\d}{\text{d}s}p^a$
 and internal
ones $\frac{\d }{\text{d}s}p^k$. 

First we evaluate  equation for external momentum evolution:
$$     
\frac{\d }{\text{d}s}p^a(z^a,[z^k])=\frac{1}{2}v^{\b}S^{\g\d }     
R^a_{\;\;\b\g\d}\;.\tag{18}
$$
Dividing summation over indices into the one over external indices and
the one over internal indices  we have:
$$
\aligned
\frac{\d }{\text{d}s}p^a &      
=\frac{1}{2}v^bS^{cd}R^a_{\;\;bcd}+\frac{1}{2}v^kS^{cd}R^a_{\;\;kcd}     
+v^bS^{ck}R^a_{\;\;bck}+v^lS^{ck}R^a_{\;\;lck}+\\     
&\qquad+ v^bS^{kl}R^a_{\;\;bkl}+v^kS^{lm}v^lR^a_{\;\;klm}\;.
\endaligned
$$
Now we can use calculated elements of curvature tensor obtaining formula:  
$$  
\frac{\d }{\text{d}s}p^a=\frac{1}{2}v^bS^{cd}R^a_{\;\;bcd}     
+\frac{1}{2}v^{l}S^{c}_{\;l}     
g^{ad}(\frac{1}{4}\frac{1}{f^4}\partial_d     
f^2\partial_cf^2-     
\frac{1}{2}\frac{1}{f^2}\partial_d\partial_cf^2+     
\frac{1}{2}\frac{1}{f^2}\G^b_{dc}\partial_bf^2)\;,
$$
that gives us:
$$
\frac{\d }{\text{d}s}p^a(z^a,[z^k])=
\frac{1}{2}v^bS^{cd}R^a_{\;\;bcd}\;.
\tag{19}
$$ 
Thus the equation of external momentum evolution does not show any
influence of assumed extendness of the particles.   

Now we can repeat the same procedure for equation of internal
components of momentum:
$$     
\aligned     
\frac{\d }{\text{d}s}p^k(z^a,[z^k])&=\frac{1}{2}v^\b S^{\g\d}     
R^a_{\;\;\b\g\d}     
=v^aS^{bl}R^k_{\;\;abl}+\frac{1}{2}v^lS^{mn}R^k_{\;\;lmn}=\\      
&=\frac{1}{2}v^lS^{mn}R^k_{\;\;lmn}     
-v^aS^{bk}(\frac{1}{4}\frac{1}{f^4}\partial_a     
f^2\partial_bf^2-     
\frac{1}{2}\frac{1}{f^2}\partial_b\partial_af^2+     
\frac{1}{2}\frac{1}{f^2}\G^c_{ba}\partial_cf^2)\;, 
\endaligned \tag{20}
$$ 
and we obtain formula:
$$
\frac{\d }{\text{d}s}p^k(z^a,[z^k])
=-v^aS^{bk}(\frac{1}{4}\frac{1}{f^4}\partial_a     
f^2\partial_bf^2-     
\frac{1}{2}\frac{1}{f^2}\partial_b\partial_af^2+     
\frac{1}{2}\frac{1}{f^2}\G^c_{ba}\partial_cf^2)\;.\tag{21} 
$$     
The second of equations of motion (7) remains unchanged:
$$\frac{\d S^{\k\la}}{\text{d}s}=2p^{{[}\k}v^{\la {]}}\;, \tag{22}$$  
thus conditions imposed on metric do not change the structure of this
equation. 

Internal components of momentum $p^k$ correspond to generalised charge
of the particles, thus the law of the charge conservation requires
$\delta p^k / \text{d}s=0$.  In 5-dimensional space-time
with one additional internal dimension we have identically $S^{kl}=0$,
i.e. mixed components of spin tensor vanish. So we have then
$\delta p^4/\text{d}s=0$   that corresponds to
conservation of electric charge in dipole approximation.

\head 5.  Multipole approximation\endhead

Now we want  to consider equations of motion in better than dipole 
approximation. Equations of motion are given by expression
(4) and (5). First we will  estimate multipole moments using
assumed  smallness of characteristic length of additional dimensions. 

The $2^n$-pole moment reads:      
$$I^{\k_1\cdots\k_n\la\mu}(z(s))=\int_{T_z{\eusm M}}X^{\k_1}
\cdots X^{\k_n}      
\hat{T}^{\la\mu}(z,x)\,\text{D}X\;.\tag{23}$$      
According to our previous assumption the metric tensor  
$g_{\a\b}$ is a block-diagonal matrix hence  its determinant is 
given by a product of external metric tensor determinant $g_E$ and
internal metric tensor determinant $g_I$:
$$\text{det}\;g_{\a\b}=\text{det}\;g_{ab}\cdot\text{det}\;g_{kl}= 
g_E\cdot g_I= (-1)^{(k-1)}f^{2(k-1)}g_E\cdot h_I\;.\tag{24}$$     
We have assumed that the whole internal space is the orbit  of symmetry  
group action hence the energy-momentum skeleton
 $\hat{T}^{\k\la}$ is constant on the orbit, i.e.:
$$\hat{T}(z,X^a,X^k)=\hat{T}(z,X^a)\;,\tag{25}$$     
Using this property we  can separate integral in expression (23)
defining $2^n$-pole moment: 
$$ 
\aligned 
I^{\k_1\cdots\k_n\la\mu}(z(s))=&f^{(k-1)}(z^a)\left(\int_{T_z{\eusm M}_E} 
\text{D}X_E\sqrt{-g_E(z^a)}X^{\k_E^{(1)}} 
\cdots X^{\k_E^{(l)}}\hat{T}^{\la\mu}\right)\cdot\\     
\cdot &\left(\int_{T_z{\eusm M}_I} \text{D}X_I\sqrt{h_I([z^k])}
X^{\k_I^{(l+1)}}\cdots X^{\k_I^{(n)}}\right)\;.
\endaligned\tag{26} 
$$

According to the kind of components $\la\mu$ we will have different 
expression for multipole moments. Let us deal with the cases:

{
\leftskip 0.5in
\item{$\bullet$} $\la,\mu$ are  external components $a, b$,  
\item{$\bullet$} $\la,\mu$ are mixed components $a,k$, 
\item{$\bullet$} $\la,\mu$ are internal components. 
\par
}

 Thus in the first case we have simple formula:    
$$
\aligned
I^{c\cdots dk\cdots l ab}(z^a, [z^k])=&f^{(k-1)}(z^a)
\left(\int_{T_z{\eusm M}_E} 
\text{D}X_E \sqrt{-g_E(z^a)} X^c\cdots X^d \hat{T}^{ab} \right)\cdot\\
\cdot&\left(\int_{T_z{\eusm M}_E} \text{D}X_I\sqrt{h_I([z^k])}
X^k\cdots X^l\right)\;,
\endaligned\tag{27}$$
This gives that the external moment  $I_E^{c\cdots dab}$ is
additionally 
multiplied  
by the constant
\linebreak $\int \text{D}X_I\sqrt{h_I}X^k\cdots X^l$ and a conformal factor 
$f^{(k-1)}(z^a)$.

In the other cases we do not have so unique interpretation of
obtained multipole terms and we have to consider what is the
interpretation of the integral:
$$\int_{T_z{\eusm M}_E} \text{D}X_E \sqrt{-g_E} X^{\k_E^{(1)}}
\cdots X^{\k_E^{(l)}}\hat{T}^{kl}\tag{28}$$ 
appearing in the expression on multipole moments::
$$
\aligned
I^{\k_1\cdots\k_nkl}(z(s))=&f^{(k-1)}(z^a)\left(\int_{T_z{\eusm M}_E} 
\text{D}X_E \sqrt{-g_E} X^{\k_E^{(1)}}\cdots  
X^{\k_E^{(l)}}\hat{T}^{kl}\right)\cdot\\
\cdot&\left(\int_{T_z{\eusm M}_I}\text{D}X_I\sqrt{h_I}     
X^{\k_I^{(l+1)}} \cdots X^{\k_I^{(n)}}\right)\;.
\endaligned
\tag{29}$$
Therefore we have to consider properties of internal components
of the energy-momentum skeleton $\hat{T}^{kl}$.  For 
energy-momentum tensor it would be energy
density, but we have subtracted from it some quantities to obtain
energy-momentum skeleton containing  less information.

We can write the second integral in expression (29)  in the form :
$$
\int_{T_z{\eusm M}_I}\text{D}X_I\sqrt{h_I(z^a)} X^{k_1}\cdots X^{k_n}=
 \sqrt{h_I(z^a)}C^{k_1\dots k_n}
\;.\tag{30}$$     
Here $C^{k_1 \dots k_n}$ are numbers dependent on internal
space geometry and the choice of the components themselves;
$g(z^a)$ is taken on group symmetry action corresponding to given
point $z^a$. We can use this quantity due to isometries of
internal space: $\text{\it\$}g_I=0$ 
on given orbit.

Now let us try to estimate multipole moments in case the test
body is point-like in external dimensions. Then  external moments
vanish: $I_E=0$ and we
could look for pure influence of existence of internal dimensions on
equations of motion.

Only moments with internal first $n$ indices: $m_1\cdots m_n$  
 have influence on equations of motion and they read:
$$I^{m_1\cdots m_n \la\mu}(z(s))=\int_{T_z{\eusm M}_E} \text{D}X_E \sqrt{-g_E}\hat{T}^{\la\mu}\int_{T_z{\eusm M}_I}
 \text{D}X_I\sqrt{g_I}     
X^{m_1}\cdots X^{m_n}\;.\tag{31}$$     
Because the body is point-like in external dimensions  we can write
the energy-momentum skeleton $\hat{T}^{\la\mu}$ as  distribution around
area $X^a=0$:
$$\hat{T}^{\la\mu}=M\d^{m-k}(X^a)\;,\tag{32}$$
i.e. $\hat{T}^{\la\mu}$ is concentrated only at the point
 $X^a=0$ and some compact area of internal space. The quantity
  $M$ is a dimensional constant.
Hence we have integral:    
$$\int_{T_z{\eusm M}_I} \text{D}X_E \hat{T}^{\la\mu}=M\;,\tag{33}$$     
and the form of multipole moments in that case is following:
$$I^{k_1\cdots k_n \la\mu}(z(s))= M\sqrt{-g(z)}C^{k_1\cdots k_n}\;.\tag{34}$$ 
As we see that moments are functions only of external coordinates.

\subhead 5.1  Momentum evolution\endsubhead 

\bigskip
We will determine now equation of motion for momentum in two cases: for external 
and internal components.

\bigskip
\item{\bf (a)} Equation of motion for external components of momentum.

\bigskip
We put into equation of momentum evolution (4) calculated  form of
multipole moments (34) and we have the following equation:
$$ 
\aligned
\frac{\d}{\text{d}s}p_a(z^a,[z^k])=&  \frac{1}{2}v^bS^{cd}R_{abcd}+\\
&+\sum_{n=2}^N\frac{1}{n!} MR^{n+k-1}C^{k_1\cdots k_n \la\mu}\nabla_{a} 
g_{\la\mu,k_1\cdots k_n}(z^a,[z^k]) \;.
\endaligned\tag{35}     
$$
From the definition of the covariant derivative it follows that it could be written in terms of the partial derivatives and Christoffel symbols:
$$\align \nabla_a g_{\la\mu,k_1\cdots k_n}=&\partial_a g_{\la\mu,k_1 
\cdots k_n}+\\   
&-\G_{a\la}^\nu g_{\nu\mu ,k_1\cdots k_n}-\dots-\G_{ak_n}^\nu g_{\la\nu,k_1\cdots \mu} 
\endalign 
$$   
We have calculated general form of Christoffel symbols according to our assumptions on the properties of the space-time and  we have that $\G_{ak}^c=0\;$ and
$\;\G^m_{ak}=\frac{1}{2}g^{mn}\partial_ag_{kn}\; ,\;$ and we can write
formula (35) as following:
$$ 
\aligned 
\frac{\d}{\text{d}s}p_a(z^a,[z^k])&=\frac{1}{2}v^bS^{cd}R_{abcd}+ 
+\sum_{n=2}^N\frac{1}{n!} MR^{n+k-1}C^{k_1\cdots k_n \la\mu}\\
&\cdot\left(\vphantom{\frac12}\partial_{a}g_{\la\mu,k_1\cdots k_n}\right. 
-\dots
\left.-\frac{1}{2}g^{mn}(\partial_ag_{k_n n}) 
g_{\la\mu,k_1\cdots m}(z^a,[z^k])\right)\;.  
\endaligned\tag{36} 
$$    
That gives us equation of external momentum evolution along the
centre-of-mass world line in  $2^n$-pole approximation.

\item{\bf (b)}   Equation of motion for internal momentum components

\bigskip
We repeat the same procedure as in the point (a) and the equation on
internal momentum evolution reads:
$$ 
\aligned 
\frac{\d}{\text{d}s}p_m(z^a,[z^k])&= 
v^aS^{\;\;b}_m\left(\frac{1}{4}\frac{1}{f^4}\partial_a     
f^2\partial_bf^2-     
\frac{1}{2}\frac{1}{f^2}\partial_b\partial_af^2+     
\frac{1}{2}\frac{1}{f^2}\G^c_{ba}\partial_cf^2\right)+\\     
&+\sum_{n=2}^N\frac{1}{n!} MR^{n+k-1}C^{k_1\cdots k_n \la\mu}\nabla_{m} 
g_{\la\mu,k_1\cdots k_n}(z^a,[z^k])\; .
\endaligned\tag{37} 
$$

In general choice of coordinate system this derivative does not vanish as we could expect in standard point-like Kaluza-Klein theories.

\subhead 5.2 Angular momentum evolution \endsubhead  

\bigskip
Now let us consider equation for angular momentum.
Equation of motion are of the following form:
$$\frac{\d}{\text{d}s}S^{\k\la}(z^a,[z^k])=2p^{{[}\k}v^{\la {]}}+      
\sum_{n=1}^{N-1}\frac{1}{n!}g^{\sigma[\k}      
I^{\la]\rho_1\cdots\rho_n\mu\nu}g_{\{\sigma\nu ,\mu\}\rho_1\cdots\rho_n}
(z^a,[z^k]) \;.\tag{38}
$$   
We will evaluate this equation in three cases of indices $\k$, $\la$.

\item{\bf (a)} Equations of external components $S^{ab}$. 

\bigskip
In that case $\k,\la=a,b$ and we have additional terms:
$$g^{c[a}I^{b]r_1\cdots r_n\la\mu}g_{\{c l ,k\}r_1\cdots r_n}\;.\tag{37}$$   
We can expand expression in brackets such that 
 $g_{\{cl,k\}}=g_{cl,k}- g_{lk,c}+g_{kc,l}\;$. 
Thus the only term different from zero is $g_{lk ,c r_1\cdots r_n}$. 
We also have that $2^n$-moment with mixed first n indices vanishes: $I^{br_1\cdots r_n\mu\nu}=0\,$,
hence all additional terms
disappear and we are left with the equation:
$$\frac{\d}{\text{d}s}S^{ab}=2p^{[a}v^{b]}\;.\tag{40}$$ 
Here we do not have higher than dipole terms due to our assumption that 
we deal with
particles being  point-like in four standard dimensions and extended in
compactified additional dimensions.
That extendness gives contribution in external space only to intrinsic
angular momentum and as spin effects appear only in the dipole term we 
do not have the higher ones.

\item{\bf (b)} Equations of mixed components $S^{am}$. 

\bigskip
Here we have additional terms as the following:
$$ 
g^{ca}I^{mr_1\cdots r_n \la\mu}g_{\{cl ,k\} r_1\cdots r_n} 
-g^{nm}I^{a r_1\cdots r_n \la\mu}g_{\{nl ,k\} r_1\cdots r_n}\tag{41} 
$$ 
and $ I^{ar_1\cdots r_n\mu\nu}=0$, 
hence we have angular momentum equation in the form: 
$$   
\frac{\d }{\text{d}s}S^{am}=-p^{k}v^{a}-\sum_{n=1}^{N-1}\frac{1}{n!}g^{ca} 
I^{m r_1\cdots r_n \la\mu} g_{\{cl ,k\}r_1\cdots r_n}(z^a,[z^k]) \;.\tag{42}
$$   

\item{\bf (c)} Equations of mixed components $S^{nm}$ 

\bigskip
In that case we just  have following equation: 
$$\frac{\d}{\text{d}s} S^{nm}=-\sum_{n=1}^{N-1}\frac{1}{n!}g^{pn} 
I^{mr_1\cdots r_n \la\mu}   
g_{\{ \la\mu,p \}r_1\cdots r_n}(z^a,[z^k]) \;,  \tag{43}
$$   
that is rather of general form.

\bigskip
All right hand sides of obtained equations (36), (37), (40), (42),
(43) give us deviation from geodesic motion. In the other case momentum
would be parallelly propagated:
$$ \frac{\d}{\text{d}s}p^\k=0\;,$$
as well as angular momentum:
$$\frac{\d}{\text{d}s}S^{\k\la}=0\;.$$

\bigskip
\head 6. Conclusions \endhead

We have derived equations of motion for particles extended in additional 
dimensions characteristic for Kaluza-Klein type theories. There are 
extra terms containing multipole moments coming form extendness of 
the bodies. Thus we have deviation from geodesic motion.

We can estimate this deviation considering the inner space being in
the simplest form  of the $(k{-}1)$-dimensional  sphere of the radius $R$ 
parametrised
by $k{-}1$ angles $\phi, \theta,\dots, \chi$. Then the coefficients 
$C^{k_1\cdots k_n}$ in Eq. (34) for multipole moments are given by:
$$
\aligned
C^{k_1\cdots k_n}&=
\int_0^{2\pi}R\text{d}\phi\int_0^{\pi}R\text{d}\theta\;\dots 
\int_0^{\pi}R\text{d}\chi\;(R^i\phi^i)(R^j\theta^j)   
\cdots (R^l\chi^l)=\\   
&=\frac{2^{j+1}\pi^{n+k-1}}{(i+1)(j+1)\cdots(l+1)}R^{n+k-1}\;,
\endaligned \tag{44}
$$  
where $i+j+\dots+l=n$.

Thus we see that multipole moments are of order $R^{n+k-1}$.
The radius of internal space is of 
the Planck scale hence additional terms in equation on momentum and 
angular momentum evolution give only slight contribution to 
equation of motion unless the space-time is quite regular.
\bigskip

\head 7. Acknowledgments      \endhead
The finacial support of the Polish Scientific Committee under the grant
2 P03B 017 12, contract number PB 1371/P03/97/12, and Polish Research Project
BST-561/T  are gratefuly 
acknowledged.

\enddocument